\documentclass[11pt]{article}
 \usepackage{lineno}
  \usepackage[dvipdf]{graphicx}
  \usepackage{color}
  \usepackage{amssymb}
  \usepackage{natbib}
  \usepackage[labelfont=bf,labelsep=period]{caption}

 \newcommand{\beq}{\begin{equation}}
 \newcommand{\eeq}{\end{equation}}
 \newcommand{\beqa}{\begin{eqnarray}}
 \newcommand{\eeqa}{\end{eqnarray}}
 \newcommand{\beqan}{\begin{eqnarray*}}   %no number
 \newcommand{\eeqan}{\end{eqnarray*}}
 \newcommand{\beqn}{\begin{eqnarray*}}    %no number
 \newcommand{\eeqn}{\end{eqnarray*}}

\begin{document}

%% ------------------------------------------------------------------------ %%
%
%  TITLE
%
%% ------------------------------------------------------------------------ %%

%\title{Solar XUV and ENA-driven water loss from early Venus' steam atmosphere}
\begin{center}
\textbf{\Large Solar XUV and ENA-driven water loss from early Venus' steam atmosphere}
\end{center}

%\maketitle

%% ------------------------------------------------------------------------ %%
%
%  AUTHORS AND AFFILIATIONS
%
%% ------------------------------------------------------------------------ %%

%Use \author{\altaffilmark{}} and \altaffiltext{}

% \altaffilmark will produce footnote;
% matching \altaffiltext will appear at bottom of page.

%\authors
\vspace{3mm}\noindent
H.~I.~M. Lichtenegger$^1$, K.~G. Kislyakova$^1$, P.~Odert$^1$, N.~V. Erkaev$^{2,3}$, H. Lammer$^1$, H.~Gr\"{o}ller$^4$,
C.~P. Johnstone$^5$, L.~Elkins-Tanton$^6$, L. Tu$^5$, M.~G\"{u}del$^5$, M. Holmstr\"{o}m$^7$

%\altaffiltext{1}{Space Research Institute, Austrian Academy of Sciences, Schmiedlstr. 6, A-8042, Graz, Austria.}
%\altaffiltext{2}{Institute of Computational Modelling, SB RAS, 660036 Krasnoyarsk, Russian Federation.}
%\altaffiltext{3}{Siberian Federal University, 660041 Krasnoyarsk, Russian Federation.}
%\altaffiltext{4}{Department of Planetary Sciences, University of Arizona, Tucson, AZ 85721, USA.}
%\altaffiltext{5}{University of Vienna, Department of Astrophysics, T\"{u}rkenschanzstra{\ss}e 17, A-1180 Vienna, Austria.}
%\altaffiltext{6}{School of Earth \& Space Exploration (SESE), Arizona State University, Tempe, Arizona, USA.}
%altaffiltext{7}{Swedish Institute of Space Physics, Box 812, SE-98128 Kiruna, Sweden.}

\vspace{3mm}\noindent
\footnotesize
$^1$ Space Research Institute, Austrian Academy of Sciences, Schmiedlstr. 6, A-8042, Graz, Austria,
$^2$ Institute of Computational Modelling, SB RAS, 660036 Krasnoyarsk, Russian Federation,
$^3$ Siberian Federal University, 660041 Krasnoyarsk, Russian Federation,
$^4$ Department of Planetary Sciences, University of Arizona, Tucson, AZ 85721, USA,
$^5$ University of Vienna, Department of Astrophysics, T\"{u}rkenschanzstra{\ss}e 17, A-1180 Vienna, Austria,
$^6$ School of Earth \& Space Exploration (SESE), Arizona State University, Tempe, Arizona, USA,
$^7$ Swedish Institute of Space Physics, Box 812, SE-98128 Kiruna, Sweden
\normalsize

%% ------------------------------------------------------------------------ %%
%
%  KEYPOINTS
%
%% ------------------------------------------------------------------------ %%

% Key points are 1 to 3 points that the author provides,
% that are 100 characters or less, that are ultimately published
% with the article.
%% for example:
% \keypoints{\item Here is the first keypoint. what happens if it is a
% long keypoint, like this one. We want to see this wrap please.
% \item This is the second.
% \item And here is the third keypoint
% }

%\keypoints{(Type in Key Points Here)}

%% Keypoints will print underneath the abstract.

%% ------------------------------------------------------------------------ %%
%
%  ABSTRACT
%
%% ------------------------------------------------------------------------ %%

% >> Do NOT include any \begin...\end commands within
% >> the body of the abstract.

\rule{\textwidth}{1pt}

\textbf{Abstract}
%\begin{abstract}
The influence of the hydrogen hydrodynamic upper atmosphere escape, driven by the solar soft X-ray and extreme ultraviolet
radiation (XUV) flux, on an expected magma ocean outgassed steam atmosphere of early Venus is studied. By assuming that
the young Sun was either a weak or moderate active young G star, we estimated the water loss from a hydrogen dominated
thermosphere due to the absorption of the solar XUV flux and the precipitation of solar wind produced energetic hydrogen
atoms (ENAs). The production of ENAs and their interaction with the hydrodynamic extended upper atmosphere, including
collision-related feedback processes, have been calculated by means of Monte Carlo models. ENAs that collide in the upper
atmosphere deposit their energy and heat the surrounding gas mainly above the main XUV energy deposition layer. It is
shown that precipitating ENAs modify the thermal structure of the upper atmosphere, but the enhancement of the thermal
escape rates caused by these energetic hydrogen atoms is negligible. Our results also indicate that the majority of
oxygen arising from dissociated H$_2$O molecules is left behind during the first 100 Myr. It is thus suggested that
the main part of the remaining oxygen has been absorbed by crustal oxidation.

\rule{\textwidth}{1pt}
%\end{abstract}

%% ------------------------------------------------------------------------ %%
%
%  BEGIN ARTICLE
%
%% ------------------------------------------------------------------------ %%

% The body of the article must start with a \begin{article} command
%
% \end{article} must follow the references section, before the figures
%  and tables.

%\begin{article}
\newpage
\section{Introduction}
%---------------------

Based on a 1-D hydrodynamic upper atmosphere model, \cite{Chassefiere:1996b,Chassefiere:1997} studied the loss of
hydrogen from a hot, H$_2$O-rich thermosphere on early Venus that was exposed to a factor of 5 higher XUV flux
compared to that of today's solar value. In addition, \cite{Chassefiere:1996a} studied the hydrodynamic escape
of oxygen that remained from dissociated H$_2$O in primitive atmospheres of early Venus and Mars for XUV fluxes
up to 20 times stronger than that of the present solar value. For removing a huge amount of water, including heavier
dissociation products such as oxygen, it was shown in \cite{Chassefiere:1996b,Chassefiere:1997} that the ratio of
the XUV flux to the solar wind strength is of prime importance. While the solar wind regulates the escape flux
from outside due to the production of energetic neutral atoms (ENAs) via charge exchange with solar wind protons,
the XUV flux acts from the inside by supplying atmospheric atoms with the energy they require to be lifted
up to the exobase level. \cite{Chassefiere:1996b} found that the XUV flux is deposited mainly in the lower
thermosphere, while the ENA flux which is directed toward the planet is absorbed at an atmospheric layer below
the exobase where ENAs also contribute to thermospheric heating.

These pioneering studies related to ENA heating suggested that under the assumed atmospheric and solar parameters,
about 75\% of the escape energy is supplied by ENAs and that any planetary magnetic field that pushes the planetary
obstacle up to altitudes higher than $\sim$3 $R_{\rm pl}$ would inhibit the additional heating effect of the ENAs.
However, at the time when this work was performed, it was thought that the XUV radiation for early Venus was not
higher than probably $\sim$5--20 times today's value and that the density of the primitive solar wind
could have been higher by a factor of 500 or even more. In addition, the solar wind velocity was considered to be
approximately similar to the present-day value of about 400 km s$^{-1}$.

Today a lot more data from multiwavelength X-ray and UV observations from various satellites exist which indicate
that the XUV flux was considerably higher than a factor of 5 or 20 as assumed in \cite{Chassefiere:1996a,Chassefiere:1996b,Chassefiere:1997}.
Observations of young main sequence stars that confirm the slowdown of their rotation due to activity and angular
momentum loss \citep{Johnstone_et_al:2015b} suggest that the X-ray and extreme ultraviolet (XUV)
emissions of the young Sun might have been much higher than the present solar value \citep{Ribas_et_al:2005,
Guedel:2007, Tu_et_al:2015}.
%Earlier observations of \cite{Ribas_et_al:2005} and \cite{Claire_et_al:2012}
%of a few selected solar proxies suggest that the average XUV flux of the young Sun at an age of about 100 Myr was
%enhanced by a factor of $\sim$100 compared to today's solar value.

Recently, \cite{Tu_et_al:2015} studied the rotation related activity evolution of many solar like stars and showed
that a spread of initial conditions leads to a wide distribution of possible X-ray luminosities in the age range of
20-170 Myr, before rotational convergence and therefore also X-ray luminosity convergence sets in. Since this stellar
age range is crucial for the evolution of young planetary atmospheres, the early planetary evolution histories are
mainly determined by the initial rotation of the planet's host star. According to the study of \citet{Tu_et_al:2015},
young solar-like G stars may be expected to exhibit an XUV flux of $\sim$30-500 times that of the present Sun during
the first $\approx$100 Myr.

Moreover, the primitive solar wind was also probably less dense during the first hundred million years after the
origin of the solar system \citep{Johnstone_et_al:2015a, Johnstone_et_al:2015b} than assumed by
\cite{Chassefiere:1996b,Chassefiere:1997}. Astrosphere stellar mass loss observation results obtained during the last
decade with the Hubble Space Telescope (e.g., \cite{Wood_et_al:2005}) together with observationally constrained
rotational evolution models \citep{Johnstone_et_al:2015a,Johnstone_et_al:2015b} can now be used to derive wind mass
loss rates and they predict stellar wind densities for the first $\sim$100 Myr not much larger than that of today's
Sun. Therefore, the wind density of the young Sun at the orbit of Venus was most likely not much different than today
so that a value of a few tens per cm$^{3}$ is expected at 0.7 AU.

In the present study, we investigate the contribution of an enhanced solar radiation and of stellar wind produced ENAs
to the heating and escape of the hydrogen-dominated thermosphere of early Venus when it is exposed to different solar
XUV flux values that could have been emitted from the young Sun during the activity saturation phase. In section 2, we
describe and justify the model input parameters, the XUV absorbtion and hydrodynamic upper atmosphere model, the solar
wind upper atmosphere interaction and ENA production model and the model for studying the energy deposition of ENAs in
the upper atmosphere. Furthermore, we present the results of the ENA-induced additional thermospheric heating process
connected to atmospheric escape. In section 3 we discuss the consequences of our results related to the loss of water from
early Venus.

\section{Model description}
%--------------------------
To investigate the solar induced XUV and ENA heating contribution to a hydrogen dominated upper atmosphere at early
Venus, we assume that the hydrogen originates from the dissociation of H$_2$O molecules of a steam atmosphere that
was degassed from the planet's interior during the solidification of a magma ocean (e.g., \cite{Albarede&Blichert-Toft:2007,
Elkins-Tanton:2012, Lebrun_et_al:2013, Hamano_et_al:2013}).

\subsection{Production of Venus' initial H$_2$O inventory and of a dense CO$_2$ atmosphere}
%------------------------------------------------------------------------------------------
Theoretical model results related to the formation of terrestrial planets suggest that Venus originated as a result of
giant impacts between large planetary embryos and most likely possessed a globally molten magma ocean (e.g., \cite{Elkins-Tanton:2012}).

According to \cite{Gillmann_et_al:2009} and the terrestrial planet formation studies of \cite{Morbidelli_et_al:2000} and
\cite{Raymond_et_al:2006}, the equivalent of up to two terrestrial water oceans (TO) may have been delivered to early Venus
during $\sim$10--35 Myr, while up to a few tens of TO may have been brought by a few large planetary embryos later during
35-170 Myr. However, it should be noted that in the studies of \cite{Morbidelli_et_al:2000}, \cite{Raymond_et_al:2006} and
\cite{Gillmann_et_al:2009}, the loss of H$_2$O from these large planetary embryos was not taken into account. Differentiated
large planetary embryos will also form magma oceans so that a significant fraction of the catastrophically outgassed
H$_2$O inventory released from the embryo's solidifying mantle can also be lost \citep{Erkaev_et_al:2014,Erkaev_et_al:2015,
Schiefer:2014, Maindl_et_al:2015}, before the growing planet finishes accretion.
\cite{Schiefer:2014} showed that a Mars-size planetary embryo that outgasses a steam atmosphere with a surface pressure of
250 bar ($\sim$1 TO) will lose this atmosphere by hydrodynamic escape at 0.7 AU in $\sim 10^{5}$ years, caused by an XUV
activity $\sim$100 times higher compared to that of today's Sun. Planets orbiting less active young solar-like stars
most likely evolve wetter compared to planets orbiting more active stars. From these studies, one can therefore assume
that only a fraction of the initial water and volatile inventory will remain in and on the protoplanet after the planet's
accretion period is finished.

The equilibrium degassing models used in the formation of dense H$_2$O and CO$_2$ atmospheres on early Earth show that for all magma ocean depths under $\sim$2000 km and all initial carbon contents of those oceans less than $\sim$200 ppm, the resulting CO$_2$ atmosphere pressure is less than the current Venusian $\sim$90 bars of CO$_2$. Modeling efforts focusing on Earth's terminal magma ocean from the Moon-forming impact indicate that melting the entire mantle is possible but unlikely \cite{Canup:2004}. Since these Earth models apply equally to Venus, we consider magma oceans deeper than $\sim$2000 km on Venus to be unlikely. How then did Venus acquire its more massive current CO$_2$ atmosphere? Venus accreted from more carbon-rich material compared to early Earth or the effects of serial magma oceans were additive and resulted in a more massive atmosphere or additional carbon was added through later events.

The first option, in which Venus accreted from more carbon-rich materials, is entirely possible. There are both chondritic and achondritic meteorites with 1,000 ppm or more of carbon  \citep{Jarosewich:1990}. Many of these more carbon-rich meteorites carry larger quantities of water. The addition of more water and more carbon to the initial atmosphere significantly slows the cooling time of the magma ocean and thus renders the atmosphere more vulnerable to stripping by the active young Sun \citep{Hamano_et_al:2013}. This stripping is capable of carrying away both water and carbon and thus may not result in a lasting massive atmosphere.
Serial magma oceans may have had a similar effect. If a fraction of the preexisting atmosphere is retained during subsequent accretionary impacts \citep{Genda&Abe:2003}, then the young Venus may have been developing an increasingly thick atmosphere as it grew. This thicker atmosphere would have the same vulnerability to stripping. Finally, and possibly most likely, carbon could have been added to the Venusian atmosphere through later processes.

Volcanism during the subsequent evolution of the planet would have added further carbon to the atmosphere. The Earth's current CO$_2$ budget in crustal rocks is about 6$\times 10^{21}$ moles \citep{Zhang&Zindler:1993}. Assuming that this current carbon reservoir on Earth is the result of volcanism (plate tectonics having removed earlier deposits of carbon to the interior), it may be a rough proxy for carbon degassing on Venus, despite their different tectonic processes. On Venus, the equivalent in atmosphere pressure of 6$\times 10^{21}$ moles of CO$_2$ is about 50 bars. Including the effects of stripping, this additional carbon is insufficient to explain the current density of Venus' atmosphere.

It should be noted, however, that carbon is also added through the tail of accretion to young Venus by the devolatilization
of impactors \citep{OKeefe&Ahrens:1977}. Again the Earth can be used as an analog. Matching the measured highly siderophile
elements in Earth's mantle requires approximately 2$\times 10 ^{22}$ kg of material with bulk chondritic composition to have
been added to the Earth after the Moon-forming impact \citep{Bottke_et_al:2010}. If this material contained 1000 ppm of carbon,
and all the carbon went into the Venusian atmosphere as CO$_2$, additional 14 bars would have been be added to the atmosphere.
Some chondritic meteorites do have as much as 30 times more carbon, 3 wt\% or more. Because the late
accretion of the Earth is thought to be dominated by just one or two large impactors \citep{Bottke_et_al:2010}, it is possible
that the majority of Venus' carbon atmosphere was contributed by the tail of accretion.

Because the energy and size of late accretionary impacts on early Venus are unknown, in this paper we consider a partially
molten mantle with a 2000 km-deep magma ocean. The outgassing from these magma ocean depths with 0.1 wt\% H$_2$O and 0.02
wt\% CO$_2$ contents can provide a steam atmospheres with partial surface pressures of 458 bar H$_2$O and 101 bar CO$_2$.

According to \cite{Lebrun_et_al:2013} and \cite{Hamano_et_al:2013}, who studied the thermal evolution of an early magma
ocean in interaction with a catastrophically outgassed steam atmosphere at early Venus, H$_2$O vapor would begin to
condense into liquid water after $\ge$10 Myr. Because Venus' orbital distance of $\sim$0.72 AU locates the planet on a
border zone where steam atmospheres may never reach condensation conditions, it is not clear if H$_2$O was sometimes
present in liquid form at the surface of Venus or if it remained always in vapor form \citep{Hamano_et_al:2013}.
However, an extreme water loss from early Venus during the magma ocean period might explain the present planet's dry interior
and the fate of the remnant oxygen \citep{Gillmann_et_al:2009, Hamano_et_al:2013, Lammer_et_al:2011, Lammer_et_al:2013}.
One should also note that for the surface temperatures of $\sim$500 K, which are expected during the ``mush'' stage
\citep{Lebrun_et_al:2013}, one can also expect water vapor mixing ratios at the mesopause level near to 1 \citep{Kasting:1988}.
For that reason H$_2$O would have continued to escape effectively, even if there are periods in which there was liquid
water on the planet's early surface.

\subsection{Stellar wind and XUV radiation of the young Sun}
%-----------------------------------------------------------
Stellar magnetic activity induced X-rays/EUV radiation declines in time and the precise evolutionary path depends on the
rotational evolution \citep{Tu_et_al:2015}. This study suggests that during the activity saturation phase, slow-, average-,
and fast-rotating young solar-like stars emit an XUV flux of $\sim$30, $\sim$ 100, and more than $\sim$500 times higher
than that of the present Sun. Because the rotation and related activity of young Sun-like stars converge after the first
Gyr, it is unknown which activity evolution track the Sun followed. Therefore, we assume in the following two different
XUV fluxes at Venus orbit, a 30 times and a 100 times higher value compared to the present Sun.

In order to estimate the wind properties that correspond to the 30x and 100x present XUV fluxes, we use the wind model
developed by \cite{Johnstone_et_al:2015a,Johnstone_et_al:2015b}. We estimate that these XUV fluxes correspond
approximately to rotation rates of 6~$\Omega_\odot$ and 12~$\Omega_\odot$ for the 30x and 100x present XUV fluxes,
respectively, assuming the current solar rotation rate to be $\Omega_\odot=2.67 \times 10^{-6}$~rad~s$^{-1}$. We consider
here only the slow component of the solar wind. The calculations are based on a 1D hydrodynamic wind model run using the
\emph{Versatile Advection Code} developed by \cite{Toth:1996} and assume a polytropic equation of state with a spatially
varying polytropic index. In order to get the wind speed at 1 AU of $\sim$400~km~s$^{-1}$ in our model, we assume a base
temperature of the wind of 1.8~MK. To calculate the wind speeds for winds of the more active Sun, we assume that the wind
temperature scales linearly with coronal temperature (i.e. model~A in \cite{Johnstone_et_al:2015a}). At the orbit of Venus,
we find wind speeds of 890~km~s$^{-1}$ and 1140~km~s$^{-1}$ for the 30x and 100x XUV flux cases, respectively. For the wind
temperature, we obtain values of $3.9 \times 10^5$~MK and $5.8 \times 10^5$~MK for the 30x and 100x XUV flux cases, respectively.
\cite{Johnstone_et_al:2015b} derived a scaling law for the wind mass loss rate based on fitting a rotational evolution model
to the observational constraints; they estimated that the stellar mass loss rate, $\dot{M}_\star$, scales with stellar radius,
$R_\star$, rotation, $\Omega_\star$, and mass, $M_\star$, as $\dot{M}_\star \propto R_\star^2 \Omega_\star^{1.33} M_\star^{-3.36}$.
This implies $\dot{M}_\star$ values of $1.2 \times 10^{-13}$ and $2.8 \times 10^{-13}$~M$_\odot$~yr$^{-1}$ and therefore
proton densities at the orbit of Venus of 35 and 63~cm$^{-3}$, for the 30x and 100x XUV flux cases, respectively.

In this study, we consider only the contribution from the quiet solar wind and not from time variable components such as coronal mass ejections (CMEs). Based on the correlation between flares and CMEs observed on the current Sun and the high flare rates of active stars, it has been suggested that the winds of highly active stars could be dominated by CMEs \citep{Aarnio_et_al:2012,Drake_et_al:2013,Osten_Wolk:2015}. Strong CMEs would compress the magnetospheres of planets \citep{Khodachenko_et_al:2007} and should cause an increase in the ENA production rate \citep{Lammer_et_al:2007,Kislyakova_et_al:2013} during the passage of the CME. However, strong CME activity on active stars has not been found observationally (e.g. \citealp{Leitzinger_et_al:2014}) and the existence of such CME winds is currently very speculative.

\subsection{XUV absorption and hydrodynamic upper atmosphere model}
\label{section:XUV}
%------------------------------------------------------------------
For the upper atmosphere parameters, we follow the assumptions by \cite{Kasting&Pollack:1983, Chassefiere:1996b}
and \cite{Gillmann_et_al:2009} that the high XUV flux of the young Sun would dissociate most H$_2$ and H$_2$O molecules
near the homopause level so that the upper atmosphere is dominated by hydrogen atoms. As shown by \cite{Marcq:2012},
during periods of magma ocean related hot surface temperatures, the tropopause location in an overlaying steam atmosphere
can be raised at an Earth or Venus-like planet from its present altitude of $\sim$30--40 km up to altitudes of
$\sim$300--550 km. Because the studied hydrogen atoms originate from a magma-ocean related degassed steam atmosphere
with a hot surface temperature, we assume the lower boundary of our simulation domain near the mesopause-homopause level
to be $R_0=R_{\rm pl}+300$ km, with a gas temperature $T_0=250$ K and a number density $n_0 = 5 \times 10^{12}$ cm$^{-3}$
(e.g., \cite{Tian_et_al:2005}.

To study the XUV-heated upper atmosphere structure and thermal escape rates of the hydrogen atoms, we apply a 1-D hydrodynamic
model with energy absorption, which is described in detail by \cite{Erkaev_et_al:2013, Erkaev_et_al:2014, Erkaev_et_al:2015}
and \cite{Lammer_et_al:2013, Lammer_et_al:2014}. The model is considered to be valid until the Knudsen number, which is the
ratio between the mean free path and the scale height, reaches 0.1 \citep{Volkov_et_al:2011}. The model solves the following
system of the hydrodynamic equations for mass,
\begin{equation}
\frac{\partial \rho R^2}{\partial t} + \frac{\partial \rho v R^2}{\partial R}= 0,
\end{equation}
momentum,
\begin{eqnarray}
\frac{\partial \rho v R^2}{\partial t} + \frac{\partial \left[ R^2 (\rho v^2+P)\right]}{\partial R} =\rho g R^2 + 2P R,
\end{eqnarray}
and energy conservation
\begin{eqnarray}
\frac{\partial R^2\left[\frac{\rho v^2}{2}+\frac{P}{(\gamma-1)}\right]}{\partial t}
+\frac{\partial v R^2\left[\frac{\rho v^2}{2}+\frac{\gamma P}{(\gamma - 1)}\right]}{\partial R}=\nonumber\\
\rho v R^2 g + Q R^2.
\end{eqnarray}
Here $R$ is the radial distance from the center of the planet, $\rho, P, T, and v$ are the mass density, pressure,
temperature and velocity of the outward flowing bulk atmosphere, respectively. Further, $\gamma$ is the adiabatic index,
$g$ the gravitational acceleration and $Q$ is the heating rate consisting of two parts
\begin{eqnarray}
Q = Q_{\rm XUV} + Q_{\rm ENA},
\end{eqnarray}
where $Q_{\rm XUV}$ is the heating rate related to the XUV absorption, and $Q_{\rm ENA}$ is the additional heating
source due to the precipitating ENA particles. To establish the $Q_{\rm XUV}$ function, we follow the approach of
\cite{Erkaev_et_al:2013, Erkaev_et_al:2014}, where $Q_{\rm XUV}$ is given by
\begin{eqnarray}
Q_{\rm XUV} = \eta\sigma n I_{\rm XUV} \frac{1}{4\pi}\int_0^{\pi/2+\arccos(1/r)}{J(r,\theta)
2\pi\sin(\theta)d\theta}.
\end{eqnarray}
Here $r$ is the normalized radial distance $r = R/R_0$, with $R_0$ being the lower boundary radius, $\eta$ is the
heating efficiency for a hydrogen atmosphere, which can be assumed as 15 \% according to \cite{Shematovich_et_al:2014},
$I_{\rm XUV}$ is the XUV flux at the upper boundary at 0.7 AU (see Table 2), $\sigma$ is the cross section of the XUV
absorbtion, and $J(r,\theta)$ is a dimensionless function describing the variation of the XUV flux due to the atmospheric
absorption,
\begin{eqnarray}
J(r,\theta) = \exp\left [-\int_r^{\tilde R^*}{a\tilde n(\xi)
\left(\xi^2 - r^2\sin(\theta)\right)^{-1/2}\xi d\xi}\right]. \label{J}
\end{eqnarray}
In this equation, $R^*$ is the upper boundary radius, $\tilde n = n/n_0$, where $n_0$ is the density at the lower boundary,
and $a$ is a dimensionless constant parameter given by $a = \sigma n_0 R_0$. Eq. (\ref{J}) describes the XUV flux
intensity as a function of the spherical coordinates $r$ and $\theta$. In the particular case when $\theta$ = 0,
this formula simplifies to an expression similar to that of \cite{Murray-Clay_et_al:2009}. The additional heating
source $Q_{\rm ENA}$ is calculated from the kinetic model of charge exchange interaction between the hydrogen atoms
in the exosphere and the incoming stellar wind protons.

\subsection{Exosphere-stellar wind interaction and ENA production modeling}
%--------------------------------------------------------------------------
We use a Direct Simulation Monte Carlo (DSMC) code to model the interaction between the stellar wind and the
hydrogen-dominated upper atmosphere. The code includes neutral hydrogen atoms, $H_{\rm pl}$, and hydrogen ions,
$H^+_{\rm pl}$, which include both ions of planetary origin and solar wind protons, $H_{\rm sw}^+$. These particles
can charge-exchange and produce ENAs, $H_{\rm ENA}$, which conserve the energy of stellar wind protons, following the
reaction
\begin{equation}\label{Eq:chargeExchange}
 H_{\rm pl} + H^{+}_{\rm sw} \rightarrow H_{\rm pl}^{+} + H_{\rm ENA}.
\end{equation}
As a result of the interaction between the stellar wind and the XUV-radiated and expanded upper atmosphere, a huge
hydrogen corona including ENAs is formed around the planet. In addition to charge exchange, the planetary atmosphere is
ionized by the stellar wind via electron impact ionization and stellar radiation. The intensity of the interaction
depends on the atmospheric parameters, stellar XUV radiation intensity, stellar wind flux and on the presence and
strength of the intrinsic magnetic field. In this study, we assumed a nonmagnetized early Venus to model the most
intensive possible interaction. The applied code is described in more details in \cite{Kislyakova_et_al:2013,Kislyakova_et_al:2014}.

The main processes and forces included for an exospheric atom are the following:
collision with a UV photon which determines the velocity-dependent radiation pressure;
charge exchange with a stellar wind proton;
elastic collision with another hydrogen atom;
ionization by stellar photons or wind electrons;
and gravity of the star and planet, centrifugal, Coriolis and tidal forces.

The parameters used for the simulations are summarized in Table~\ref{Tab:parameter}. We have applied the DSMC code assuming an
XUV flux 30 and 100 times higher than that of the present Sun, corresponding to slow and moderate rotators, in order to test
whether heating by ENAs can provide a significant contribution to the total loss.

\begin{table*}%[b]
\begin{minipage}{\textwidth}
  \caption{Inner boundary conditions of a hydrogen dominated upper atmosphere and stellar wind and radiation
  input parameters applied to the DSMC atmosphere interaction model.}
  $
  \begin{array}{p{0.5\linewidth}llll}
    \hline
    \noalign{\smallskip}
    Name &  \rm Symbol & 30\;{\rm XUV} & 100\;{\rm XUV} \\
    \noalign{\smallskip}
    \hline
    \noalign{\smallskip}
Inner boundary radius, $[R_{\rm pl}]$     		& R_{\rm ib}			& 4.95 					& 4.57    \\
Inner boundary temperature, [K]					& T_{\rm ib}			& 1790 					& 3470    \\
Inner boundary density, $[\rm cm^{-3}]$ 		& n_{\rm ib}			& 2.00 \times 10^{6} 	& 3.12 \times 10^{6}   \\
Inner boundary velocity, [km/s] 				& v_{\rm ib}	       	& 4.98		            &  7.5   \\
Obstacle standoff distance, $[R_{\rm pl}]$		& R_{\rm s} 	    	& 5.0 					& 4.69     \\
Obstacle width, $[R_{\rm pl}]$					& R_{\rm t} 	  		& 5.0 					& 4.69     \\
Photoionization rate, $[\rm s^{-1}]$ 			& \tau_{\mathrm{pi}}	& 9.0\times 10^{-6} 	& 3.0\times 10^{-5}    \\
Electron impact ionization rate, [s$^{-1}]$	    & \tau_{\mathrm{ei}} 	& 5.0\times 10^{-7} 	& 7.5\times 10^{-7}   \\
Solar wind density, $[\rm cm^{-3}]$				& n_{\rm sw}     		& 3.46\times 10^1 		& 6.29\times 10^1   \\
Solar wind velocity, [km/s]			 	    	& u_{\rm sw}     		& 892 \times 10^3 		& 1138 \times 10^3    \\
Solar wind temperature, [K]				    	& T_{\rm sw}     		& 3.9 \times 10^5 		& 5.8 \times 10^5    \\
    \noalign{\smallskip}
    \hline
  \end{array}
  $
\label{Tab:parameter}
\end{minipage}
\end{table*}

In this study we use a velocity-dependent absorption rate of photons \citep{Kislyakova_et_al:2014} which has the same shape as the stellar Ly $\alpha$ line  so that the strongest acceleration of neutral hydrogen atoms occurs in the velocity domain between approximately $-100\le v_x \le 100$~km/s. The absorption rate is defined as $\beta_\mathrm{abs}=\int \sigma(\lambda) \Phi(\lambda) \mathrm{d}\lambda$, where $\sigma(\lambda)$ is the absorption cross-section and $\Phi(\lambda)$ is the stellar Ly $\alpha$ spectrum \citep{Meier:1995}. The absorption cross-section is $\sigma(\lambda)=\int \psi(v) \sigma_N(\lambda') \mathrm{d}v$, where $\psi(v)$ is the normalized atomic velocity distribution and $\sigma_N(\lambda')$ is the natural absorption cross-section of an atom traveling with velocity $v$ for which $\lambda'=\lambda(1-v/c)$. To infer an average Ly $\alpha$ profile of the young Sun, we use the reconstructed line profiles of three young solar analog stars from \cite{Wood_et_al:2005}. Since there are no data available for a Sun-like star with an age of about 100 Myr, we use the intrinsic profiles of three older G-stars (HD 59967, 200 Myr; $\chi^1$ Ori, 300 Myr; $\kappa^1$ Cet, 650 Myr) and scale their integrated Ly $\alpha$ fluxes to a common age of 100 Myr using eq. 2 of \cite{Ribas_et_al:2005}. The line profiles are then averaged and scaled to the orbit of Venus.

Figure~1 %\ref{Fig:PhScatter}
shows the obtained Ly $\alpha$ absorption rate dependent on the radial velocity of the neutral hydrogen atoms. Absorption of a stellar Ly $\alpha$ photon is followed by a subsequent radiation of a photon in a random direction. In average, this process leads to substantial acceleration of neutral hydrogen atoms and deformation of the hydrogen corona surrounding the planet (Figure~2). %\ref{Fig:corona}).

%------------------------------------------------------------------
\begin{figure}[t]
%\stepcounter{figure}
%\setcounter{figure}{0}
\label{Fig:PhScatter}
\begin{center}
\includegraphics[width=0.6\textwidth]{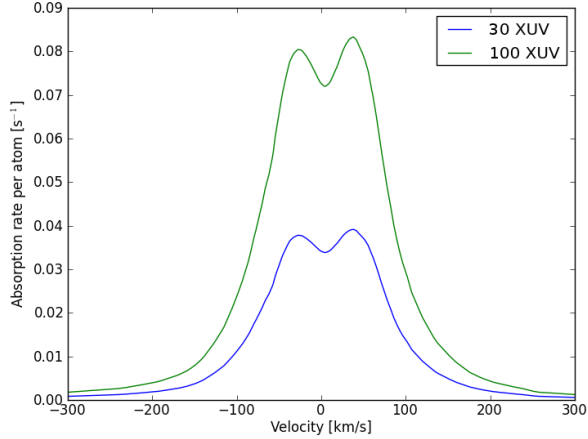}
\caption{Ly $\alpha$ absorption rate dependent on the radial velocity of hydrogen atoms for two considered cases: 30~XUV (slow rotator) and 100~XUV (medium rotator). Positive velocities denote motion toward the star. One can clearly see that the acceleration is stronger for the medium rotator and the strongest for H atoms with radial velocities $-100\le v_x \le 100$~km/s.}
\end{center}
\end{figure}
%-----------------------------------------------------------------

As an example, Figure~2 %\ref{Fig:corona}
presents the modeled corona around a young hydrogen-rich terrestrial planet with a mass and size of Venus orbiting a 100 Myr old moderate rotating young Sun emitting a 100 times higher XUV radiation than today. Hydrogen atoms are launched from the inner boundary, $R_{\rm ib}$, assuming a number density of $2.0 \times 10^{12}$~m$^{-3}$ and a temperature of 1790~K. A stellar wind with a density of $3.46\times 10^7$~m$^{-3}$, velocity of 892~km s$^{-1}$ and temperature of $3.9 \times 10^5$~K (calculated using the stellar wind model developed by \cite{Johnstone_et_al:2015a} and described in Section \ref{section:XUV}) is launched on the right side of the domain and cannot penetrate inside the ionospheric obstacle. The obstacle has a paraboloid shape \citep{Khodachenko_et_al:2012} with a width of 5.0~$R_{\rm pl}$ and the substellar point is located in 5.0~$R_{\rm pl}$ from the planet's center.
Even at this distance from the star, the thermal pressure of the atmospheric gas together with the dynamic pressure of its expansion is high enough to hold off the incoming solar wind. Above this height, the atmosphere can interact with the penetrating solar wind. We note that the choice of a narrow magnetosphere with the width equal to the distance to the subsolar point corresponds to the modeling of the solar wind interaction with a nonmagnetized body (e.g., \citealp{Baumjohann_and_Treumann:1996}). This is an extreme case of a huge hydrogen envelope surrounding a young planet.

%------------------------------------------------------------------
\begin{figure}[h!]
%\addtocounter{figure}{1}
\begin{center}
\label{Fig:corona}
\includegraphics[width=1.0\textwidth]{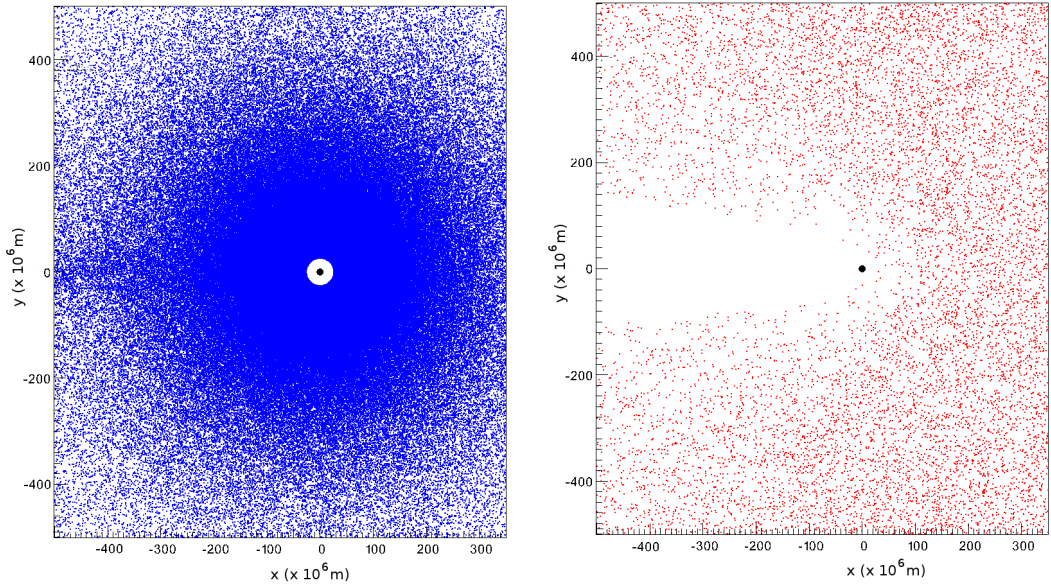}
\caption{Slice of modeled 3-D atomic H corona around early Venus according to our input parameters at 100 XUV (Table~\ref{Tab:parameter}). Blue and red dots correspond to neutral H atoms and H ions, including stellar wind protons, respectively. The black dot in the middle represents Venus, the Sun is to the right. The white area around the planet corresponds to the XUV-heated and expanded thermosphere below the inner simulation boundary of the DSMC solar wind interaction model. Stellar wind protons are launched on the right side of the domain and cannot penetrate inside the ionospheric obstacle. (left) Neutral H atoms and ENAs in the hydrogen corona. (right) Solar wind protons and ions of planetary origin. The particles from the same simulation are shown separately for convenience.}
\end{center}
\end{figure}
%-----------------------------------------------------------------

At present, Venus is a nonmagnetized planet, and despite the lack of knowledge about a possible dynamo of Venus which might have existed in the past, in this study we assume that Venus has always been nonmagnetized. Nonmagnetized planets have narrower obstacles generated by their induced magnetospheres so that the solar wind can more easily interact with planetary particles. From observations of ENAs around nonmagnetized solar system bodies it is known that their solar wind interaction region is vast (e.g., \citealp{Holmstroem_et_al:2006,Galli_et_al:2008a,Galli_et_al:2008b}), with ENAs being produced in the nose as well as in the flanks of the induced magnetosphere. Since we are mostly interested in the effect ENAs could have on the energy budget of the atmosphere, we focus on the nose interaction region, where the ENA velocities are directed mainly toward the planet. Since we do not include the deflection of the proton flux, our ENA flux is somewhat overestimated and represents an upper limit. The positions in the simulation domain and velocities of ENA metaparticles (every metaparticle represents $N$ real neutral hydrogen atoms after charge exchange) are used as an input for the upper atmosphere precipitation modeling (Section \ref{section:precipitation}).

%\vspace{4mm}
\subsection{ENA upper atmosphere precipitation modeling}
\label{section:precipitation}
%-------------------------------------------------------
Since the charge-exchange reaction (eq. \ref{Eq:chargeExchange}) does not appreciably change the energy or flow
direction of the involved particles, the newly born solar wind ENAs in the vicinity of Venus continue to move
radially away from the Sun. Therefore, the dayside of Venus is exposed to a continuous impact of ENAs interacting
with the upper atmosphere, where the energetic particles transfer energy via collisions with the atmospheric species
and thus contribute to the heating of the atmosphere. This energy deposition is simulated by means of a Monte Carlo
model which calculates the collision dominated stochastic path of the ENAs through the Venusian thermosphere.

The initial ENA flux is obtained from the ENA production model described in Section 2.4. Assuming a continuous flow
of ENAs, hydrogen atoms are launched at the dayside of Venus up to a radial distance of $\sim 60,000$ km with a
production rate according to the results of the ENA production model.
The simulation domain is divided into a number of spherical layers with a radial width of 400 km.
The atmospheric density within a layer is considered constant. The initial velocity distribution of the test particles
is chosen to mimic the corresponding distribution obtained by the ENA production model. The trajectories of the ENAs are
then followed by taking into account the elastic collisions with the hydrogen atmosphere. The energy lost by each particle
within a layer is recorded and considered as heat input at the altitude of the corresponding layer. The time step for the
motion of the particles is chosen so that the mean distance covered by the particles does not exceed the mean free path in
a layer.
%------------------------------------------------------------------
\begin{figure}[t!]
\begin{center}
\includegraphics[width=0.65\textwidth]{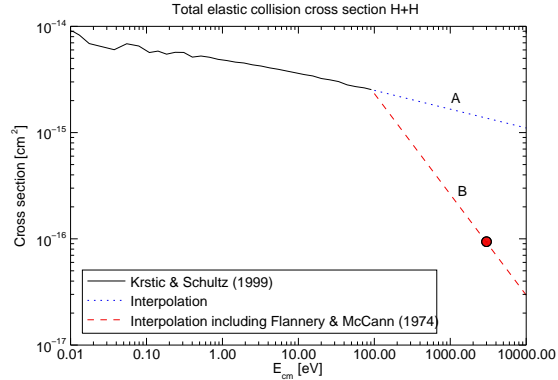}\vspace{-5mm}
\caption{Total cross section for elastic H+H collisions as a function of the center-of-mass energy $E_{\rm cm}$.
Black: total collision cross section of \cite{Krstic&Schultz:1999}; dotted blue: total collision cross section of
\cite{Krstic&Schultz:1999} extrapolated to 10$^4$ eV; dashed red: total collision cross section of \cite{Krstic&Schultz:1999}
extrapolated to 10$^4$ eV  by including the data point of \cite{Flannery&McCann:1974} (red dot on line $B$).}
\label{Fig:crossSection}
\end{center}
\end{figure}
%------------------------------------------------------------------

Concerning the interaction between the ENAs and the background atmosphere, two different assumptions were considered:
a constant H-H elastic collision cross section of $\sigma=3\times 10^{-15}$\,cm$^2$ and the total and differential collision
cross sections according to \cite{Krstic&Schultz:1999} which are given for energies between 10$^{-2}$ and 10$^2$ eV.
To our knowledge, no H-H collision cross sections for energies larger than $\sim$100 eV are available besides a single
value at $3\times 10^3$ eV published by \cite{Flannery&McCann:1974}. Since the energy of the ENAs near Venus is in the
range of several keV, we have extrapolated the data of \cite{Krstic&Schultz:1999} to 10$^4$ eV a) without including
the data point of \cite{Flannery&McCann:1974} and b) by including this value (see Figure 3). %\ref{Fig:crossSection}).
The differential elastic cross section of \cite{Krstic&Schultz:1999} shows a complete symmetry around the scattering
angle of 90$^\circ$ and the forward and backward peaks become more pronounced with increasing energy. Due to the lack of more
appropriate data we use in our calculations for the differential cross section the values of \cite{Krstic&Schultz:1999}
corresponding to 100 eV.

As can be seen from Figure 3, %\ref{Fig:crossSection},
the results of \cite{Flannery&McCann:1974} suggest a significant drop in
the elastic cross section above $\sim 100$ eV. Therefore it should be noted that the lack of data in the energy range
relevant for the ENAs hinders the correct estimation of the energy deposition rate of the ENAs and must be considered
as approximate.

After modeling the XUV absorption in the hydrogen dominated upper atmosphere, we solve the hydrodynamic equations with
the before discussed input parameters and a corresponding XUV flux during the activity saturation phase.

\subsection{H$_2$O escape}
%---------------------------------------------
Here we study the escape of an outgassed H$_2$O/CO$_2$ steam atmosphere as discussed in Section 2.1. For the initial values
we adopt the 2000~km deep magma ocean from which partial surface pressures of 458 bar H$_2$O and 101 bar CO$_2$ originate.
The H$_2$O molecules in the upper atmosphere will be dissociated by the high XUV flux of the young Sun. Therefore, the H
atoms will be the numerically dominant species in the upper thermosphere. Moreover, oxygen atoms produced by H$_2$O
dissociation should also populate the lower hydrogen dominated thermosphere and can be dragged along with the outward
flowing atomic hydrogen \citep{Zahnle&Kasting:1986, Zahnle_et_al:1990, Chassefiere:1996a, Chassefiere:1996b, Hunten_et_al:1987,
Erkaev_et_al:2014, Schiefer:2014}. On the other hand, since CO$_2$ molecules are less dissociative, a large fraction of them
might be expected to remain in molecular form. In the following calculations we therefore consider H$_2$O to be completely
dissociated into atoms, while CO$_2$ remains molecular. Under these assumptions atomic hydrogen is initially the most abundant
species ($N_\mathrm{H}/N_\mathrm{tot}=0.65$), followed by atomic oxygen ($N_\mathrm{O}/N_\mathrm{tot}=0.32$) and CO$_2$ molecules ($N_\mathrm{CO_2}/N_\mathrm{tot}=0.03$).

The escape of the outgassed atmospheric components can be described by
\begin{equation}
\dot{M}_\mathrm{H} + \dot{M}_\mathrm{O} + \dot{M}_\mathrm{CO_2} = \frac{\pi R_0^3 \eta F_\mathrm{XUV}}{G M_{\rm pl}} \equiv \dot{M},
\end{equation}
where $\dot{M}_\mathrm{H,O,CO_2}$ are the mass escape rates of H, O, and CO$_2$, $M_{\rm pl}$ is the mass of Venus, $R_0$ is the
mesopause/homopause level (see Section 2.3), $G$ is the gravitational constant, and $F_\mathrm{XUV}$ is the XUV flux of the young
Sun at the orbit of Venus; the right-hand side represents the total atmospheric mass-loss rate $\dot{M}$. It is assumed that the
absorption of the XUV radiation occurs close to $R_0$. The heating efficiency $\eta$ is taken to be 15\% (as for a hydrogen atmosphere,
see Section 2.3), though it should be noted that $\eta$ depends on the atmospheric composition and may be different for H$_2$O/CO$_2$
steam atmospheres. It should also change with time as the atmospheric composition is altered by escape. If escape is efficient, the
heavier atoms can be dragged along with the hydrogen outflow. The escape rates of the heavier species $i$ can be expressed by means
of the fractionation factor $x_i$ via $\dot{M}_i = \mu_i f_i x_i \dot{M}_\mathrm{H}$, where $\mu_i = m_i/m_\mathrm{H}$ is the atomic
mass relative to that of H and $f_i = N_i/N_\mathrm{H}$ is the mixing ratio relative to hydrogen. A simple analytic expression for
$x_i$ in a two-component atmosphere was derived by \cite{Hunten_et_al:1987} and has been applied frequently to study the escape of H$_2$O
\citep{Chassefiere:1996a, Gillmann_et_al:2009, Lammer_et_al:2011, Luger&Barnes:2015}. However, the derivation of this formula assumes
that the heavy species is a minor component and it may thus become invalid for dissociated H$_2$O atmospheres, especially if oxygen
starts to accumulate, since O is then a major component of the atmosphere. Therefore, we use an analytic formula which is not based
on this assumption to describe the escape of O \citep{Zahnle_et_al:1990}
\begin{equation}
x_\mathrm{O}=1-\frac{4\pi m_\mathrm{H} GM_{\rm pl}(m_\mathrm{O}-m_\mathrm{H})b_\mathrm{H,O}}{\dot{M}_\mathrm{H}kT_0(1+f_\mathrm{O})},
\end{equation}
where $b_{H,O}$ is the binary diffusion parameter of O in H and $T_0$ is the temperature in the lower thermosphere, taken to be 250~K
as in Section 2.3. The factor $4\pi$ stems from the fact that eq. 8 assumes that escape takes place over the whole sphere of the planet.
Note that this expression is equivalent to that of \cite{Hunten_et_al:1987} for $f_O\ll1$. For CO$_2$, we use an analytic expression
describing the escape of a minor species in the presence of two major species \citep{Zahnle_et_al:1990}
\begin{eqnarray}
x_\mathrm{CO_2} & = & \frac{\alpha \exp(\alpha/R_0)}{\alpha + \gamma - \gamma \exp(\alpha/R_0)}, \\
\alpha   & = & \frac{\dot{M}_\mathrm{H}}{4\pi m_\mathrm{H}} \left(\frac{f_\mathrm{O}(1-x_\mathrm{O})}{b_\mathrm{H,O}} + \frac{f_\mathrm{O} x_\mathrm{O}}{b_\mathrm{O,CO_2}} + \frac{1}{b_\mathrm{H,CO_2}}\right) - \nonumber \\
         &   & - \frac{G M (m_\mathrm{CO_2} - m_\mathrm{H})}{k T_0}, \nonumber \\
\gamma   & = & - \frac{\dot{M}_\mathrm{H}}{4\pi m_\mathrm{H}} \left(\frac{1}{b_\mathrm{H,CO_2}} + \frac{f_\mathrm{O}}{b_\mathrm{O,CO_2}}\right). \nonumber
\end{eqnarray}
Binary diffusion parameters of $b_\mathrm{H,O}=4.8\times10^{17}T_0^{0.75}~\mathrm{cm^{-1}~s^{-1}}$, $b_\mathrm{H,CO_2}=8.4\times10^{17}T_0^{0.6}~\mathrm{cm^{-1}~s^{-1}}$, and $b_\mathrm{O,CO_2}=7.86\times10^{16}T_0^{0.776}~\mathrm{cm^{-1}~s^{-1}}$
are taken from \cite{Zahnle&Kasting:1986}. The fractionation of the minor component CO$_2$ is affected by the presence of both H and O.
Note that $x_i$ ($i=$O, CO$_2$) cannot be negative and $x_i = 0$ means that the species does not escape, whereas $x_i \approx 1$ means that it
escapes efficiently. Eqs. 9 and 10 are analytic approximations based on several simplifications, like the assumption of subsonic and
isothermal escape \citep{Zahnle&Kasting:1986, Zahnle_et_al:1990}. However, these results are similar to those of more detailed numerical
calculations, if the masses of the heavier constituents are much larger than that of the main light species (here atomic H) and if the
fractionation factors are not too small ($x_i>1/\mu_i$). Since both O and CO$_2$ are much heavier than H and since the escape of H is
efficient under the enhanced XUV fluxes considered here, we assume that the fractionation factors and thus the escape fluxes of the heavy
constituents can be reasonably described by these analytic approximations.

\begin{table*}[b!]
\caption{Hydrodynamic loss rates without ($L_{\rm th}$) and with additional ENA heating ($L_{\rm th,ENA}$) corresponding to the
assumed XUV flux values at Venus orbit location at 0.7 AU for two activity saturated young Sun cases. $L_{\rm th,ENA}$ is
given for a constant total collision cross section (C), the laboratory measured cross section of \cite{Krstic&Schultz:1999} (A),
and the cross section of \cite{Krstic&Schultz:1999} with the result of \cite{Flannery&McCann:1974} included (B).
The XUV flux values are enhanced by the factors shown in column one and the fluxes in erg cm$^{-2}$ s$^{-1}$ correspond to
Venus' orbital location at 0.7 AU.}
\begin{center}
\begin{tabular}{l|c|c|c|c|c}
Young Sun & $I_{\rm XUV}$ &$L_{\rm th}$ & C: $L_{\rm th,ENA}$ &A: $L_{\rm th,ENA}$ & B: $L_{\rm th,ENA}$ \\
& [erg cm$^{-2}$ s$^{-1}]$ & [s$^{-1}$] &  [s$^{-1}$] &[s$^{-1}$] &  [s$^{-1}$] \\\hline
30 XUV    &  280  & 8.6 $\times 10^{31}$  & 8.6$\times 10^{31} $ &  8.7$\times 10^{31}$ & 9.5$\times 10^{31}$   \\
100 XUV   &  930  & 1.66$\times 10^{32}$  &1.68$\times 10^{32}$ & 1.72$\times 10^{32}$ & 1.82$\times 10^{32}$    \\\hline
\end{tabular}
\end{center}
\normalsize
\end{table*}

To obtain the evolution of the atmospheric inventories, we express the hydrogen mass loss rate as
\begin{equation}
\dot{M}_\mathrm{H} = \frac{\dot{M}}{1 + \mu_\mathrm{O} f_\mathrm{O} x_\mathrm{O} + \mu_\mathrm{CO_2} f_\mathrm{CO_2} x_\mathrm{CO_2}}.
\end{equation}
Since $x_i$ depends on $\dot{M}_\mathrm{H}$, we solve the above equation numerically. The escape rates for H obtained
with eq.~11 are $5.9\times 10^{31}$ and $7.5\times 10^{31}\rm\; s^{-1}$, respectively, and are slightly lower than the
escape rates from the hydrodynamic model for a pure H atmosphere (Table 2). This
is because H escape is less efficient if a fraction of its energy is needed to drag along the heavier constituents. The
mass loss rates of the heavy species are then $\dot{M}_i = \mu_i f_i x_i \dot{M}_\mathrm{H}$ if $x_i > 0$ or zero otherwise.
The evolution of the atmospheric partial masses is then obtained by numerical integration of the escape rates.

\section{Results and discussion}
%-------------------------------
%------------------------------------------------------------------
\begin{figure}[b!]
\begin{center}
\includegraphics[width=0.6\textwidth]{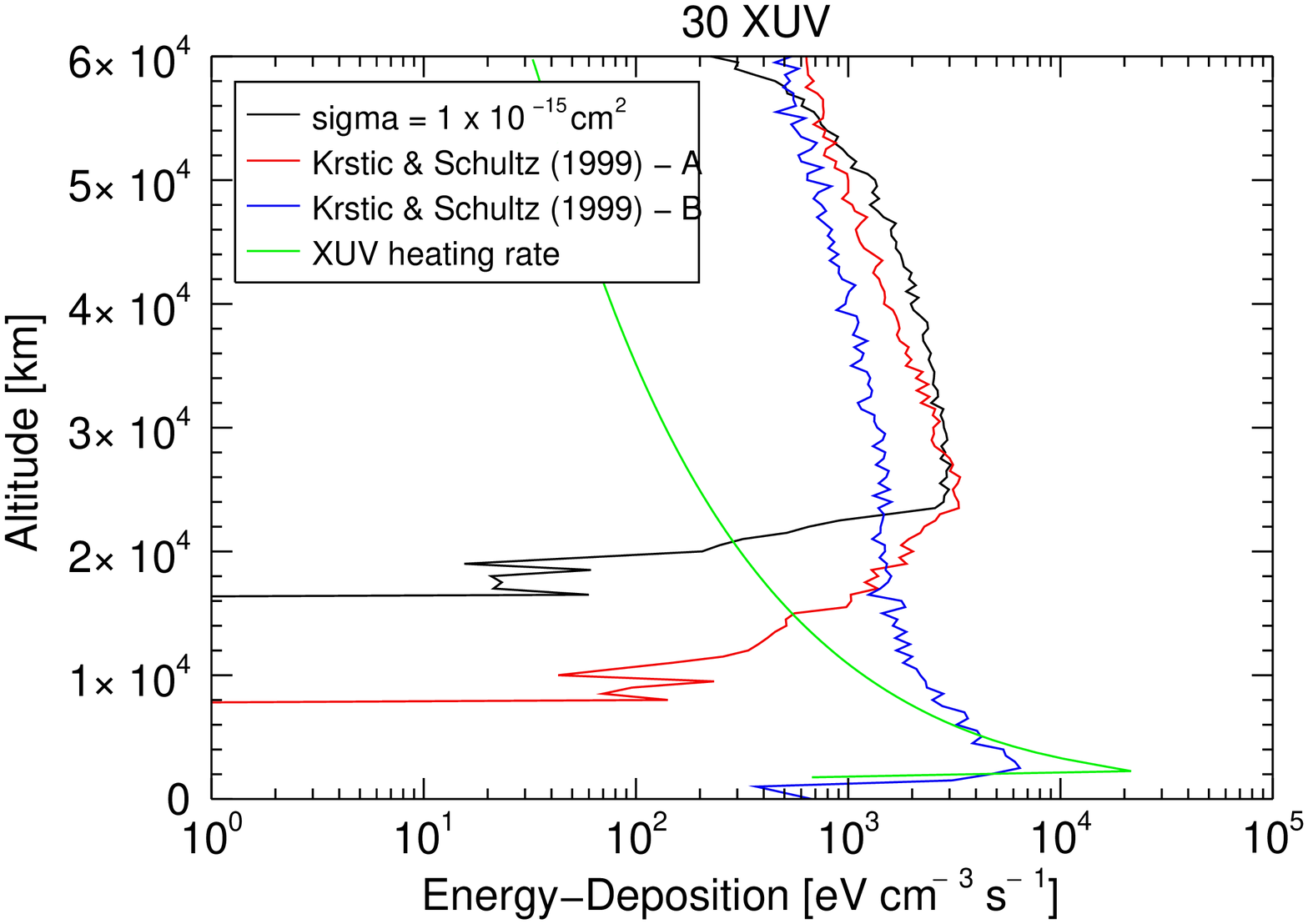}\vspace*{-4mm}
\includegraphics[width=0.6\textwidth]{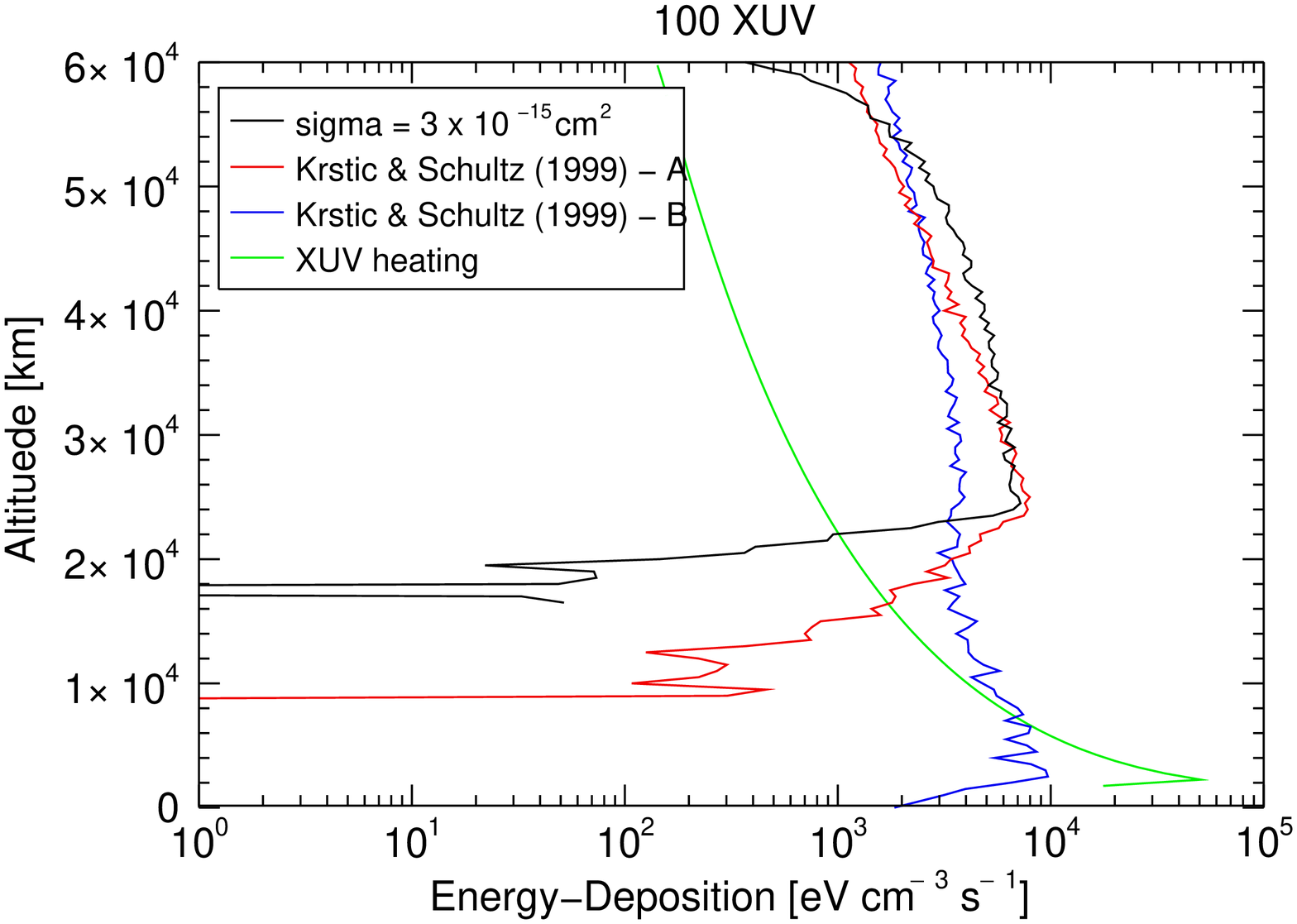}
\caption{Energy deposition profiles for the 30 and 100 XUV case (upper and lower panel, respectively) due to solar XUV flux (green line)
and due to ENAs for different collision cross sections: black: constant total collision cross section; red: total and differential
cross sections of \cite{Krstic&Schultz:1999} (line $A$ in Figure \ref{Fig:crossSection}); blue: total and differential cross sections
of \cite{Krstic&Schultz:1999} (line $B$ in Figure \ref{Fig:crossSection}).}
\end{center}
\label{Fig:EnergyDeposition}
\end{figure}
%------------------------------------------------------------------
The energy deposition rates based on three different elastic cross sections for an XUV flux of 30 and 100 times the
present solar value at 0.7 AU are displayed in Figure 4. %\ref{Fig:EnergyDeposition}.
For $\sigma=3\times 10^{-15}$ cm$^2$ as
well as for $\sigma$ varying according to \cite{Krstic&Schultz:1999} (line $A$ in Figure 3), %\ref{Fig:crossSection}),
the ENAs
quickly lose their energy and are absorbed by the background gas above $\sim 20,000$ km altitude. When the data of
\cite{Flannery&McCann:1974} (line $B$ in Figure 3) %\ref{Fig:crossSection})
are included, due to the smaller values of the total
cross section, the energetic H atoms can penetrate deeper into the atmosphere. However, the maximum energy deposition
rate is always smaller than the maximum rate caused by the solar XUV absorption.

\begin{figure}[t!]
\vspace*{-10mm}
\begin{center}
\includegraphics[width=0.6\textwidth]{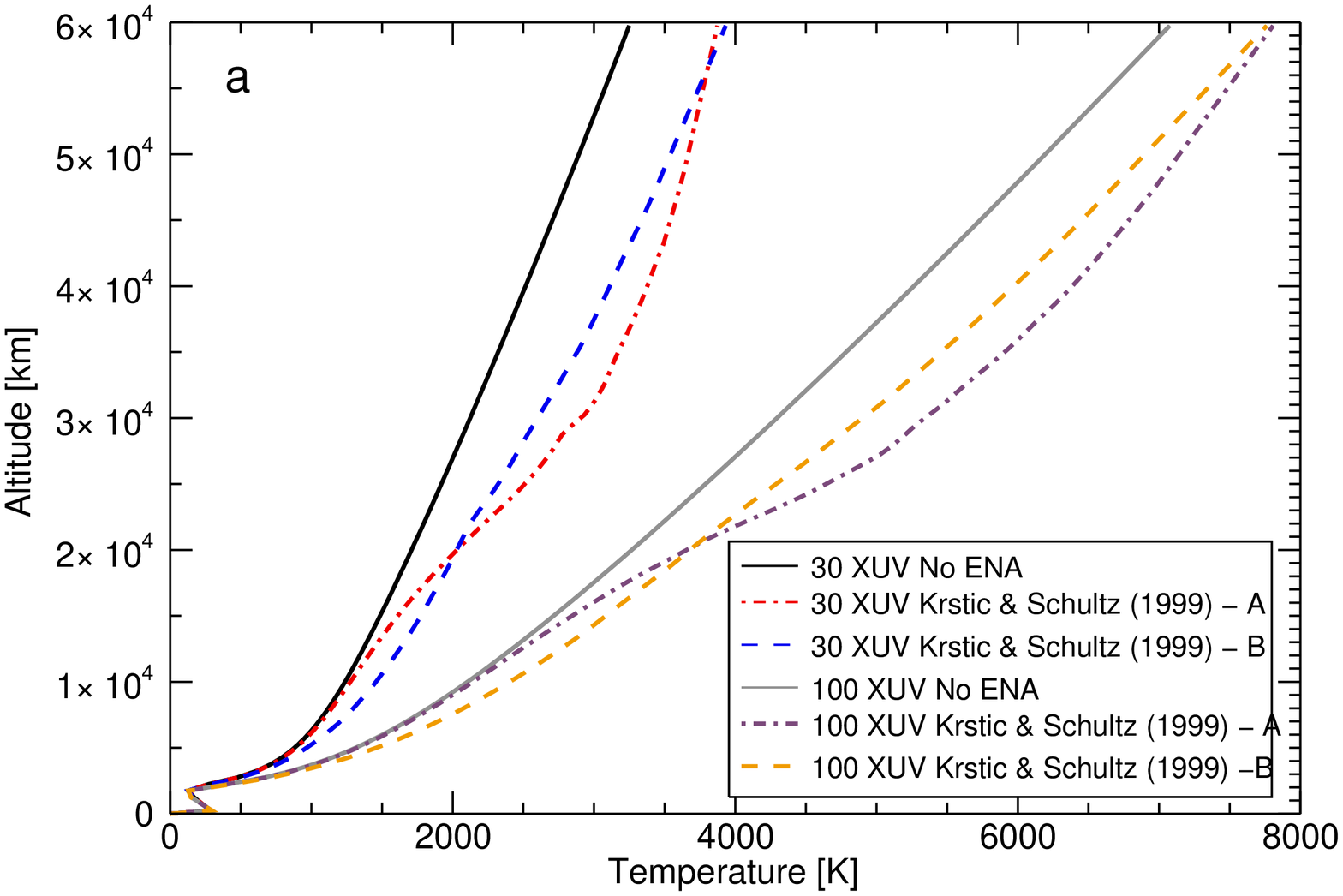}\vspace{-7mm}
\includegraphics[width=0.6\textwidth]{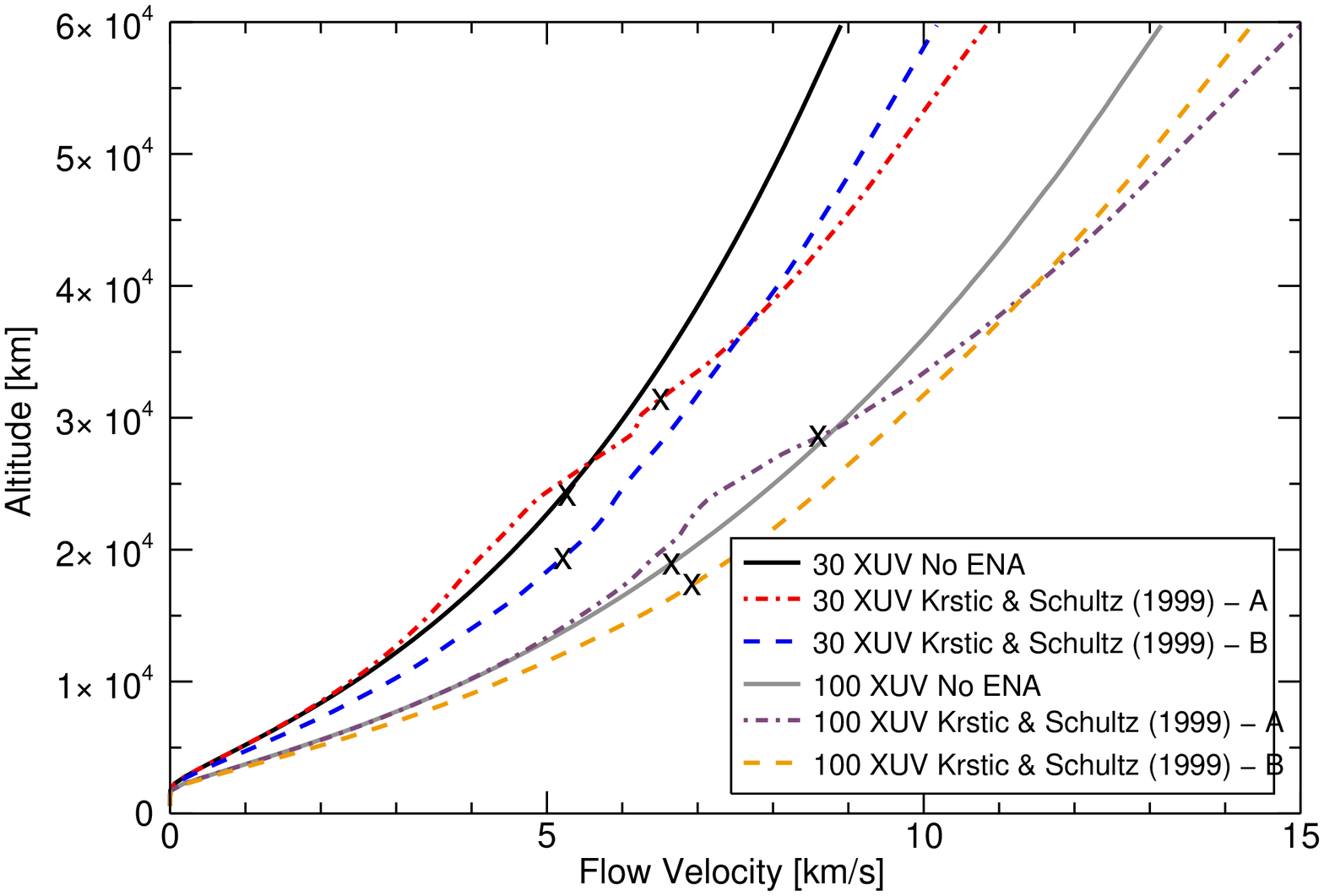}
\caption{Temperature (a) and velocity (b) altitude profiles of the hydrogen atmosphere for three different XUV fluxes.
The solid lines correspond to the case when only XUV heating is considered; the dashed lines illustrate the combined
effect of XUV and ENA heating, where Krsti\'{c} \& Schultz (1999) - A,B correspond to \cite{Krstic&Schultz:1999} extrapolated
according to $A$ and $B$ shown in Figure \ref{Fig:crossSection}. The additional dark line in (b) represents the escape
velocity at Venus and the crosses indicate the location of the sonic point.}
\label{Fig:Temp}
\end{center}
\end{figure}

The influence of the additional ENA heating on the hydrogen atmosphere profiles is shown in Figure \ref{Fig:Temp}, where the
dashed and dashed-dotted lines illustrate the effect of the ENA energy deposition on the temperature and velocity profile
of the background atmosphere.

%------------------------------------------------------------------
%-----------------------------------------------------------------

The resulting temperature and velocity profiles for XUV values of 30 and 100 times the present value are shown by the solid
lines in Figure \ref{Fig:Temp}. As can be seen, the upward flow velocity of atomic hydrogen starts to exceed the escape velocity
at an altitude of $\sim$ 20,000 (30 XUV case) and $\sim$ 15,000 km (100 XUV case), respectively.

Table 2 summarizes the escape rates $L_{\rm th}$ of atomic hydrogen obtained for the 30 times and 100 times present XUV fluxes, respectively,
without ENAs (column 3) and with ENAs taken into account (columns 4-6). These results suggest that the additional heating effect by
ENAs which originate inside the huge exosphere enhances the thermal escape rate only by a small amount. Furthermore, our results
indicate that the uncertainty in the collision cross section is of minor importance. The main reason that additional ENA heating
is less effective compared to the previous studies by \cite{Chassefiere:1996b, Chassefiere:1997} is the much lower number density
of the early solar wind at the orbital location of Venus. If the solar wind would be 1000 times denser, as assumed in these early
studies, then many more ENAs would be produced and their energy input into the thermosphere would be accordingly higher. From our
findings, we conclude that ENA heating as initially suggested by \cite{Chassefiere:1996b, Chassefiere:1997} could play an important
role for planets that are exposed to dense stellar winds in orbits closer than that of Venus. Future studies with realistic parameters
should investigate if ENA heating could enhance the thermal loss rates of terrestrial planets in habitable zones of dwarf stars or
for Venus-like planets in close orbits around their stars.

%-----------------------------------------------------------------
\begin{figure*}[b!]
\begin{center}
\includegraphics[width=0.47\textwidth]{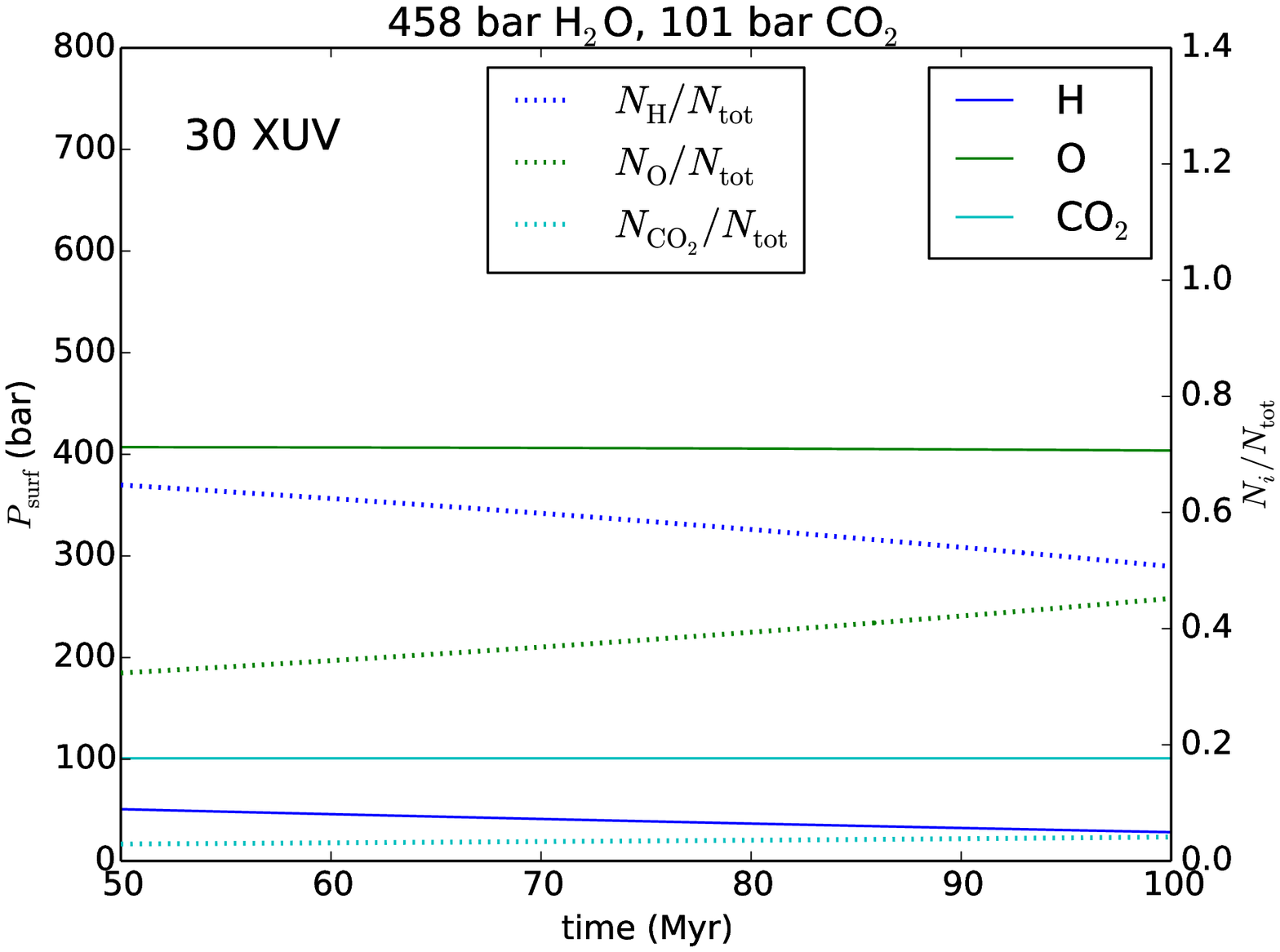}\hspace{4mm}
\includegraphics[width=0.47\columnwidth]{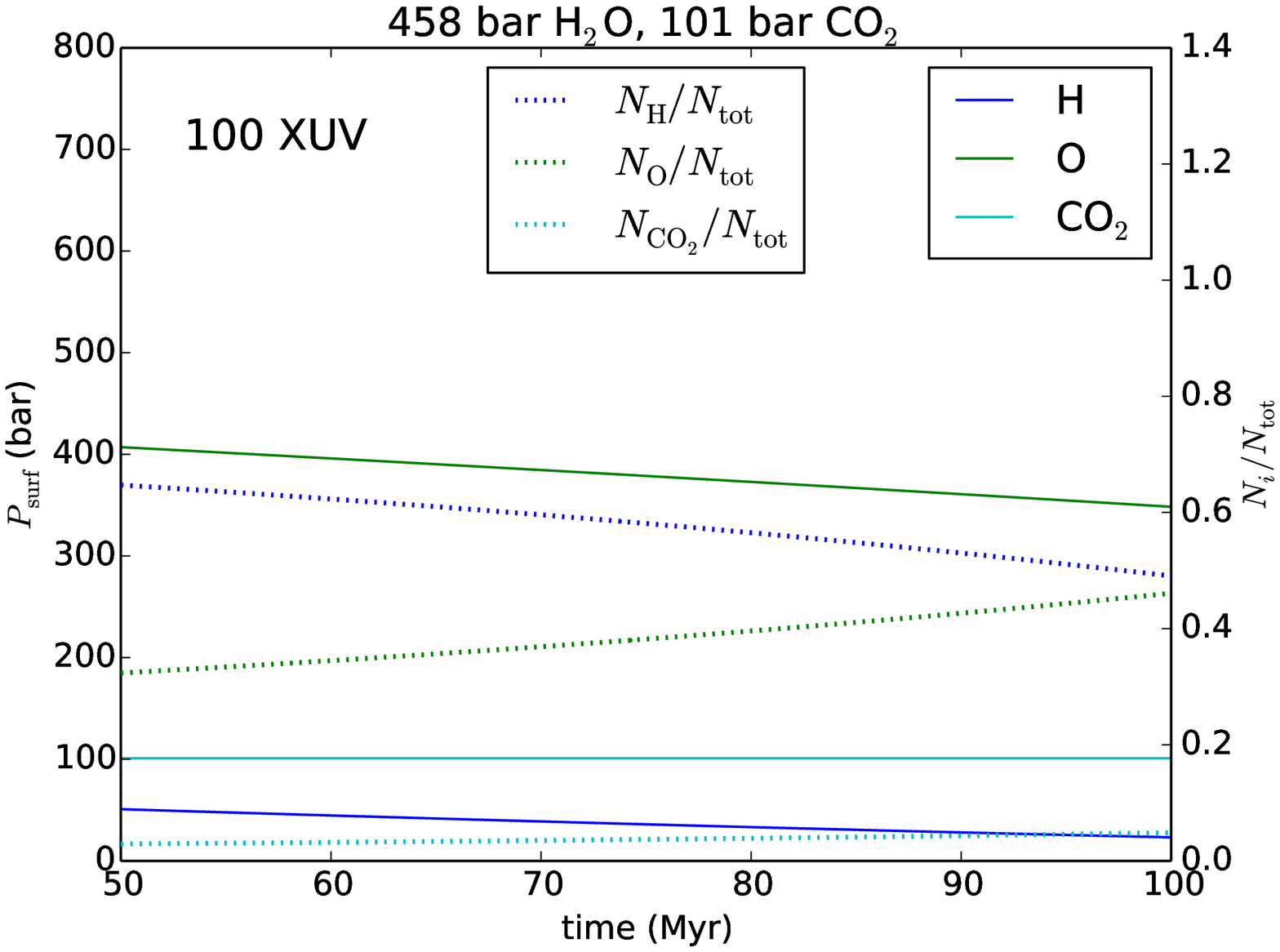}
%\includegraphics[width=0.75\columnwidth]{Fig_4c.eps}
%\hspace{2.5cm}\includegraphics[width=0.75\columnwidth]{Fig_4d.eps}
%\vspace{2.5cm}
\caption{Temporal evolution of the partial surface pressures $P_{\rm surf}$ of H, O, and CO$_2$ for a (left) slowly
and (right) moderately active young Sun with corresponding XUV enhancement factors of
30 and 100. The initial inventories of the species are determined from the outgassed atmosphere with 458~bar
H$_2$O and 101~bar CO$_2$. Dotted lines denote the mixing ratios $N_i/N_\mathrm{tot}$ of the respective species. The surface pressure evolves assuming static conditions.}
\label{Fig:SurfacePressure}
\end{center}
\end{figure*}
%-----------------------------------------------------------------

Figure \ref{Fig:SurfacePressure} illustrates the temporal evolution of the partial surface pressures of H, O, and CO$_2$ of the
outgassed atmosphere initially composed of 458~bar H$_2$O and 101~bar CO$_2$ for the cases of a weakly and moderately active
young Sun 30 and 100 times more luminous in XUV than at present.
Here we assume static conditions and do not take into account any magnetic field, non--thermal escape, or CO$_2$ dissocitation.
We also ignore the evolution of the planet's radius and the heating efficiency with changing atmospheric composition, yielding a constant total mass loss rate $\dot{M}$ during the considered time interval.
We start the calculations at an age of 50~Myr to insure
that the final accretion of Venus has been completed and follow the evolution of the atmosphere up to an age of 100~Myr, which
corresponds approximately to the end of the saturation phase of a Sun-like star, and assume that no additional material is
delivered during the escape period. Furthermore, we assume that oxygen is not removed by oxidation of the crust during this
period so that the amounts and mixing ratios change only due to atmospheric escape. It follows that H is partially removed in
both cases, while O remains almost unchanged for the weakly active case; only $\sim$15\% of its initial mass is lost in the
moderately active case (Figure~\ref{Fig:SurfacePressure}). According to \cite{Gillmann_et_al:2009}, the residual amount of O
could be absorbed into a magmatic surface, since the oxygen of three Earth oceans could be removed if there existed magma
oceans during about 100~Myr. We therefore suggest that the oxygen left behind during the main hydrodynamic phase on Venus
has been dissolved in the magma ocean and lost through oxidation.

A fraction of residual oxygen could also be lost to space via solar wind induced ion pick up \citep{Kulikov_et_al:2006}.
However, by using the solar wind parameters given in Table 1 and comparing it with the cases for the young Sun assumed
in \cite{Kulikov_et_al:2006}, it turns out that the expected oxygen ion pick up escape caused by the young solar wind
is only in the order of a few bar (see \cite{Kulikov_et_al:2006}, Figure 9) and therefore negligible compared to the
amount that should be oxidized into the magmatic surface.

After the saturation phase of the young Sun, the XUV flux remained high enough so that the light hydrogen atoms continued to
be lost by thermal escape. After $\sim$500 Myr , this will result in a water global equivalent layer of a few meters depth most
likely under the form of water vapor in the atmosphere which led to the present D/H atmospheric enhancement factor of 150.

As can be seen in Figure \ref{Fig:SurfacePressure}, CO$_2$ molecules can neither directly escape for the 30 XUV nor for the
100 XUV case, which is consistent with the present content of Venus. It should be noted, however, that due to the enhanced
XUV flux a fraction of the CO$_2$ molecules is also dissociated \citep{Tian:2009} so that part of the less massive dissociation
products may also escape, finally leading to a reduction of the initial amount of CO$_2$.
Our results are in general agreement with previous studies despite slightly different methodologies and assumptions for the young Sun's XUV emission and wind \citep{Zahnle&Kasting:1986, Chassefiere:1996a, Chassefiere:1996b, Gillmann_et_al:2009, Lammer_et_al:2011, Luger&Barnes:2015}.

If remnant H accreted from the protoplanetary nebula was mixed into the outgassed atmosphere, the results could be different.
We explore this by assuming a highly active young Sun (XUV 500 times higher than today) and add 100~bar of H to the outgassed
atmosphere. The adopted value of 100~bar corresponds to a hydrogen envelope mass fraction of $\sim10^{-4}$, which is in good
agreement with typical values found from accretion modeling (e.g. \cite{Lammer_et_al:2014}). Such an envelope would require
about 35~Myr to escape for the 500 XUV flux scenario. In the extreme case that the additional H is well mixed with the
outgassed material, even CO$_2$ molecules may escape along with the other atmospheric constituents. On the other hand, adding
H in the assumed weak and moderate activity cases slows down the escape of both H and O and more atmospheric material remains
after 100~Myr. In a more realistic scenario, however, the outgassed material may not be well mixed and the upper atmosphere is
populated predominantly by H. Then H would protect the underlying outgassed atmosphere until it is significantly depleted and
dissociation of H$_2$O and the related loss of water would start at later times. Moreover, if the accreted H envelope was too
massive, it may even reduce the outgassing rates and extend the outgassing time.

We note that the amounts of outgassed H$_2$O correspond to maximum values. If the building blocks which formed Venus had
a lower water content because they already lost a large part of it due to the high XUV emission of the young Sun and the
lower gravity of such objects, scenarios where the H$_2$O/CO$_2$ ratio of the outgassed atmosphere could be less than
assumed here are possible. Because of such potential formation scenarios, the upper possible XUV activity of the young
Sun during its earliest periods can only be estimated by studying the photochemical stability of the remaining CO$_2$
molecules. The assumption that CO$_2$ molecules remain in the upper atmosphere results in a conservative upper limit to
the atmospheric loss estimates, since CO$_2$ molecules provide an effective IR-cooling mechanism in the 15$\mu$-band and
other band emissions, which slows down the heating and the expansion of the upper atmosphere. However, if a CO$_2$-dominated
atmosphere was not maintained, its dissociation products, including other species such as N$_2$ and H$_2$, would also be lost
on much shorter timescales.

\section{Conclusion}
%-------------------
By assuming an early Venus steam atmosphere, the water loss powered by the solar XUV flux and by ENAs produced via charge exchange
with the solar wind has been studied. By means of a hydrodynamic upper atmosphere model and a Monte Carlo model, the volume heating
rates due to the absorption of the solar XUV flux and the precipitation of ENAs in the hydrogen-dominated upper atmosphere has been
calculated. It was found that in case the young Sun was either a weak or moderate active young G star, the hydrogen atoms reach escape
energies at altitudes between 15,000 and 20,000 km altitude. Although precipitating ENAs modify the thermal structure of the upper
atmosphere, they have little effect on the enhancement of the thermal escape. However, in case of a low XUV radiation but a dense
stellar wind, thermal escape may be predominantly driven by ENA heating.
It should be noted that the proton distribution in our model is obtained by assuming the protons to move along straight lines parallel to the Sun-Venus line rather than following the bent flow lines around the magnetospheric obstacle. This overestimates the number of charge-exchanged ENAs which are directed toward the atmosphere and therefore also the atmospheric heating by ENAs. However, even this overestimated ENA flux barely increases the atmospheric escape. Taking into account the flow lines would only decrease the overall energy deposition by ENAs and, therefore, not change the main conclusion.
Although we studied the conditions of a young Sun (assuming the Sun being a slow and a moderate rotator), this result is similar to the one obtained for modern Venus \citep{Shematovich_et_al:2014b}, where also the effects of the induced magnetic field on the proton flux have been taken into account.
Our results also suggest that most of the initial CO$_2$ inventory cannot be removed by the XUV fluxes less than 100x the present
solar value, while hydrogen originating from H$_2$O may escape efficiently. The remaining oxygen may be stored in a magmatic surface,
in agreement with \cite{Gillmann_et_al:2009}. A validation of this hypothesis could be obtained by collecting precise data of noble
gases in the atmosphere of Venus and by knowledge of the water content and oxidation state of Venus' surface rocks.\\

\textbf{Acknowledgements}
%------------------------
H. I. M. Lichtenegger acknowledges support from the FWF project P24247-N16,
N. V. Erkaev, M. G\"{u}del, K. G. Kislyakova,  C. P. Johnstone, and H. Lammer, acknowledge the support by the FWF NFN project S116 ``Pathways to Habitability: From Disks to Active Stars, Planets and Life'', and the related FWF NFN subprojects, S11604-N16 ``Radiation \& Wind Evolution from T Tauri Phase to ZAMS and Beyond'', and S11607-N16 ``Particle/Radiative Interactions with Upper Atmospheres of Planetary Bodies Under Extreme Stellar Conditions''. N. V. Erkaev, H. Lammer, and P. Odert acknowledges also support from the FWF project P27256-N27. N. V. Erkaev was also supported by the RFBR grant No 15-05-00879-a.
This research was conducted using resources provided by the Swedish National Infrastructure for Computing (SNIC) at the High Performance Computing Center North (HPC2N), Umea University, Sweden. The software used in this work was in part developed by the DOE-supported ASC / Alliance Center for Astrophysical Thermonuclear Flashes at the University of Chicago.

%\newpage

\bibliography{Lichtenegger-et-al_JGR_2015}

% \begin{thebibliography}%{bib_ENA}
% \bibitem
% Text
% \end{thebibliography}

%\end{article}

\end{document}